\documentclass[useAMS,usenatbib]{mn2e} \usepackage{graphicx} 
\title{Electron-positron pair production near the Galactic Centre and the 
511 keV emission line} \author[Chan]{Man Ho Chan 
\thanks{chanmh@ied.edu.hk}\\ Department of Science and Environmental 
Studies, The Hong Kong Institute of Education}

\begin{document}

\date{Accepted XXXX, Received XXXX}

\pagerange{\pageref{firstpage}--\pageref{lastpage}} \pubyear{XXXX}

\maketitle

\label{firstpage}

\begin{abstract}
Recent observations indicate that a high production rate of positrons 
(strong 511 keV line) and 
a significant amount of excess GeV gamma-ray exist in our Galactic bulge. 
The latter issue can be explained by $\sim 40$ GeV dark matter 
annihilation through $b \bar{b}$ channel while the former one remains a 
mystery. On the other hand, recent studies reveal that a large amount of 
high density gas might exist near the Galactic Centre million years ago to 
account for the young, massive stars extending from 0.04 pc - 7 pc. In 
this article, I propose a new scenario and show that the 40 GeV dark 
matter annihilation model can also explain the required positron 
production rate (511 keV line) in the bulge due to the existence 
of the high density gas cloud near the supermassive black hole long time 
ago.
\end{abstract}


\section{Introduction}
In the past few decades, a flux of 511 keV photons $\phi_{511} \sim 
10^{-3}$ ph cm$^{-2}$ s$^{-1}$ emitted in the Milky Way was reported 
\citep{Leventhal,Knodlseder}. These 
511 keV photons are supposed to originate from the positrons produced. The 
predicted production rate of positrons in the bulge and disk are 
respectively given by $\dot{N}_{e^+}=11.5^{+1.8}_{-1.44} \times 10^{42}$ 
s$^{-1}$ and $\dot{N}_{e^+}=8.1^{+1.5}_{-1.4} \times 10^{42}$ s$^{-1}$ 
\citep{Prantzos,Perets}. In particular, the bulge to disk ratio is 
abnormally high $B/D=1.42^{+0.34}_{-0.30}$ \citep{Prantzos,Perets}. The 
positrons produced in the Galactic disk can be explained by some 
mechanisms, such as novae, supernovae, pulsars and low-mass X-ray binaries 
(LMXRB) \citep{Prantzos}. However, the 
high positron production rate in the bulge remains a mystery, especially 
at the central part. Observations 
indicate that the 511 keV line comes from mainly diffuse 
sources rather than point sources \citep{Cesare}. It has been 
suggested that the positrons produced through dark matter annihilation can 
account for the required production rate \citep{Boehm}. For example, 
Ascasibar et al. (2006) show that 
dark matter annhilation can give enough positrons, and the dark matter 
mass should be smaller than 100 MeV if the annihilation cross section is 
velocity-independent. Later, based on the gamma-ray spectral 
shape, Sizun et al. (2006) further constrain the dark matter mass to 
smaller than 7.5 MeV. Although these models can give enough positrons to 
account for the 511 keV line, we do not have another independent promising 
observational evidence to support these models. Also, most of the models 
depend strongly on the dark matter density profile, annihilation 
cross-section and the dark matter annihilation channels. 

Recently, an excess of GeV gamma-ray near the Galactic Centre has been 
reported \citep{Goodenough,Hooper,Gordon,Abazajian,Daylan}. The spectrum 
obtained can be best fitted with the annihilation 
of dark matter particles with mass $m \sim 40$ GeV through $b \bar{b}$ 
channel. The required annihilation cross-section $<\sigma v>$ is about 
$(1-7) \times 10^{-26}$ cm$^3$ s$^{-1}$ \citep{Abazajian,Daylan}, which is 
consistent with the expected canonical thermal relic abundance 
cross-section $<\sigma v>=3 \times 10^{-26}$ cm$^3$ s$^{-1}$ in cosmology. 
Based 
on this result,  Boehm et al. (2014) try to use this model to explain the 
511 keV line. However, the production rate is too small to account for the 
required rate \citep{Boehm2}.

In this letter, I suggest a possible mechanism to account for both the 511 
keV line and the GeV gamma-ray excess. By assuming the existence of a 
large dense cloud near the supermassive black hole $\sim 10^6$ years ago 
and following the model that used to account for the GeV gamma-ray excess, 
the production rate of positrons can satisfy the required rate.

\section{The dense cloud near the supermassive black hole}
Recent studies indicate that a large amount of dense gas $\sim 
10^5M_{\odot}$ in the form of a disk might exist near 
the Galactic Centre ($r \le 0.4$ pc) $10^{6.5}$ years ago to account for 
the young, massive stars extending from 0.04 pc - 0.4 pc 
\citep{Wardle,Lucas,Wardle2}. The existence of the dense gas can 
overcome tidal shear in the vicinity of the supermassive black hole 
and help to explain the truncation of the stellar surface density within 
0.04 pc. Part of the cloud is converted to stars and the remainder is 
accreted onto the 
supermassive black hole \citep{Wardle2}. The accretion can last for $\sim 
10^{14}$ s, and the number density of the cloud can be as high as $n_g 
\sim 10^{10}$ cm$^{-3}$ \citep{Wardle2}. Most particles in the gas are 
electrically neutral because the temperature is not enough for ionization 
\citep{Wardle2}. Furthermore, recent studies also indicate that the 
density of gas was $n_g \sim 2\times 10^8$ cm$^{-3}$ at 1 pc from 
the Galactic Centre million years ago to overcome tidal shear 
\citep{Zadeh}. The size of such a high density region is probably 
larger than $5-7$ pc \citep{Goicoechea,Zadeh}. 

In the following, I propose that a large amount of positrons can be 
produced in the dense gas through 
pair-production mechanism ($\gamma \rightarrow e^++e^-$). If the photons 
produced by dark matter annihilation have energy greater than $2m_ec^2$, 
pair-production is possible in the field of the nucleus from the 
surrounding gas. The pair-production cross-section for photon energy 
greater than 0.1 GeV is $\sigma_{pp} \approx 9 \times 10^{-27}$ cm$^2$ 
\citep{Longair}, and it would be smaller for lower energy. 
Although this cross-section is small, the amount of positrons produced 
would still be large if the number of gas nuclei is large. 

Therefore, when a large amount of gamma-ray produced by dark matter 
annihilation is passing through this dense cloud of gas, a large amount of 
positrons can be produced. The positrons produced can also generate 
secondary photons by bremsstrahlung process. These photons produced can 
further produce high-energy positrons. As a result, cascades of 
electrons, positrons and photons would be produced. Assume that the mean 
density and the size of the dense cloud are $n_g \sim 10^8$ cm$^{-3}$ and 
$R \sim 5$ pc respectively. The optical depth of 
the electron-positron pair-production is $\tau \approx n_g \sigma_{pp} R 
\sim 13$. If we assume that the gas cloud density is falling with 
distance $n_g \propto 1/r^3$ \citep{Zadeh}, the optical depth is 
$\tau=\int_{\rm 0.4~pc}^{\rm 5~pc}n_g \sigma_{pp}dr \approx 17$. For 
$\tau=13-17$, each high-energy photon entering the cloud 
would generate $\sim 1000$ positrons in the pair-production mechanism 
\citep{Longair}. The pair-production rate is efficient when the photon 
energy is greater than 0.1 GeV. 

Since the optical depth is large, the energy of the final positrons 
produced would be $E \approx 2m_ec^2 \approx 1$ MeV (still relativistic). 
This satisfies with the strong constraints on higher energy positrons 
from in-flight annihilations ($E \le 3$ MeV) \citep{Beacom}. When the 
positrons leave the dense cloud, they would cool down mainly by 
synchrontron loss, inverse Compton scattering, bremsstrahlung loss and 
coulomb loss \citep{Longair,Storm}:
\begin{equation}
b(E)=b_s \gamma_e^2+b_i \gamma_e^2+b_bn_H \left( \frac{E}{1~\rm eV} 
\right)+b_cn_e \left[1+\frac{\ln(\gamma_e/n_e)}{75} \right],
\end{equation}
where $b_s=6.6\times 10^{-14}$ eV/s, $b_i=6.5\times 10^{-15}$ eV/s, 
$b_b=3.7 \times 10^{-16}$ eV/s, $b_c=6.1 \times 10^{-7}$ eV/s, $n_H$ is 
the number density of hydrogen atom in cm$^{-3}$, $n_e$ is 
the electron number density in cm$^{-3}$, and $\gamma_e$ is the Lorentz 
factor of a positron. Here, we assume that the magnetic field strength 
in the Galactic bulge is of the order $B \sim 10^{-5}$ G \citep{Muno}. 

The survival probability of positrons before forming positroniums is 
\citep{Prantzos}
\begin{equation}
P(E,E_f)=1-\exp \left[-n_e \int_{E_f}^{E} \frac{v(E') 
\sigma_a(E')dE'}{b(E')} \right],
\end{equation}
where $\sigma_a$ is the annihilation cross-section and $E_f$ is the final 
energy of positrons after cooling. In order to form positronium, a 
positron should cool down to lower than 100 eV. For $E=1$ MeV, $E_f=100$ 
eV and 
$n_e \sim 0.1$ cm$^{-3}$ near the Galactic Centre, only less than 1\% of 
positrons would annihilate with electrons before forming positroniums. 
Fig.~1 shows that our model satisfies the constraint of MeV photon 
spectrum observed \citep{Beacom}. A relativistic positron can cool down 
to non-relativistic in $\sim 10^{13}$ s if the injection energy 
is $\sim 1$ MeV (see Fig.~2). This means that the 
positrons produced $\sim 10^{14}$ s ago by pair-production mechanism in 
the dense gas would use $\sim 10^{13}$ s to cool down to 
non-relativistic. Then they would combine with hydrogen atoms to form 
positroniums and emit 511 keV line. Basically, since we can still observe 
the strong 511 
keV line nowadays, the cooling time scale $t_c$ should be less than the 
time of existence of the gas cloud $t_e$. Moreover, the cooling time 
scale should be longer than the timescale of disappearance of the gas 
cloud $t_a$ (the duration between the present time and the moment of the 
disappearance of the gas cloud). Based on our calculations, the condition 
$t_e>t_c$ is satisfied. However, since $t_a$ is hard to confirm, we assume 
that $t_e>t_a$ in the following discussion.

In fact, the positrons produced in the dense cloud would finally be 
trapped by the magnetic field to within 100 pc \citep{Beacom}. In other 
words, the 511 keV line signal produced by these positrons would only 
extend to 100 pc from the Galactic Centre, which seems inconsistent with 
the morphology observed ($\sim 1$ kpc in size) 
\citep{Knodlseder}. Nevertheless, the 511 photon flux from the outer part 
of Galactic bulge ($\ge 100$ pc) can be explained by the morphology of 
LMXRB ($\phi_{511} \approx 2 \times 10^{-3}$ cm$^{-2}$ s$^{-1}$ 
rad$^{-1}$) \citep{Bird,Weidenspointner,Prantzos}. Our model is going to 
account for the unexplained 
strong 511 keV line from the central part of Galactic bulge 
($\phi_{511} \approx 6 \times 10^{-3}$ cm$^{-2}$ s$^{-1}$ rad$^{-1}$).

The rate of dark matter annihilation within a radius $R$ is given by
\begin{equation}
\dot{N}_{DM}=\int_0^R \frac{\rho_{DM}^2}{m^2}<\sigma v>4 \pi r^2dr,
\end{equation}
where
\begin{equation}
\rho_{DM}=\rho_{\odot} \left( \frac{r}{r_{\odot}} \right)^{-\gamma} \left[ 
\frac{1+(r/r_s)^{\alpha}}{1+(r_{\odot}/r_s)^{\alpha}} 
\right]^{-(\beta-\gamma)/\alpha}.
\end{equation}
Following \citet{Cirelli}, we take $\rho_{\odot}=0.3$ GeV/cm$^3$, 
$r_{\odot}=8.5$ 
kpc, $r_s=20$ kpc, $\alpha=1$, $\beta=3$ and $\gamma=1.26$ (the best-fit 
value to account for the GeV excess). By using Eq.~(3) and assuming 
$<\sigma v>=3 \times 10^{-26}$ cm$^3$ s$^{-1}$, the dark matter 
annihilation rate within 5 pc is $\dot{N}_{DM}=8 \times 10^{37}$ 
s$^{-1}$. By using the spectrum calculated in \citet{Cembranos}, the total 
number of 
photons produced with energy greater than 0.1 GeV through dark matter 
annihilation is 115 per one annihilation (see Fig.~3). Assuming $\tau=13$, 
the total 
number of positrons produced in this pair-production mechanism within 
5 pc is $\sim 10^{43}$ s$^{-1}$, which is enough to account for the 511 
keV line. 

On the other hand, the diffuse source of positrons produced directly from 
dark matter annihilation can be estimated by using the positron energy 
spectrum obtained from \citet{Borriello} (see Fig.~4). The total number of 
positrons produced per one annihilation is 12. By using Eq.~(3), 
$\dot{N}_{DM} \sim 10^{39}$ for $R=1$ kpc. Therefore, the rate of diffuse
positron production within 1 kpc is about $10^{40}$ s$^{-1}$. This 
minor diffuse source of positrons requires a bit longer time to cool 
down to non-relativistic (see Fig.~2). They would finally contribute to 
the 511 keV line which extend to 1 kpc from the Galactic Centre. However, 
this minor diffuse source is negligible compared with the other available 
explanations.

The rate of change of total number of positrons in the bulge is given by
\begin{equation}
\frac{dN_{e^+}}{dt}=\dot{N}_{e^+}-N_{e^+}n_H\sigma_{eH}v,
\end{equation}
where $\sigma_{eH}$ is the cross-section of positronium production. 
Since the energy dependence of the cross-section is an exponential 
function of the positron energy, the positronium formation rate would be 
the largest for certain positron energy ($\sim 14$ eV) \citep{Boehm2}. 
As a result, a dynamical equilibrium state would be achieved when the 
positrons are cooled to around 14 eV and become positroniums. In 
equilibrium, the total number of cooled positrons in the bulge is $N_{e^+} 
\sim 10^{50}$ because $\dot{N}_{e^+} \sim 10^{43}$ s$^{-1}$ and 
$n_H\sigma_{eH}v \sim 10^{-7}$ s$^{-1}$. However, part of the dense cloud 
would eventually change to stars, and most of the remaining part would be 
captured by the supermassive black hole. Therefore, the relatively strong 
511 keV line would not last forever. When the dense cloud disappears, the
production rate would reduce to $\dot{N}_{e^+} \sim 10^{42}$ s$^{-1}$ 
(from novae, supernovae, pulsars, LMXRB, etc.). Also, the final batch of 
positrons produced from the dense cloud would diffuse away and cool down 
to form positroniums to give 511 keV line. As a result, from the 
solution of Eq.~(5), the total number of positrons would 
decrease significantly at $t_c$ after the disappearance of the dense 
cloud:
\begin{equation}
N_{e^+}=\frac{\dot{N}_{e^+}}{n_H\sigma_{eH}v} 
\left(1-e^{-n_H\sigma_{eH}vt} \right)+N_{e^+0}e^{-n_H\sigma_{eH}vt},
\end{equation}
where $N_{e^+0} \sim 10^{50}$ is the original number of positrons, and 
$t$ is the time after all the positrons produced from the dense cloud 
disappeared. When $t \gg 10^7$ s, the total number of positrons in the 
bulge would reduce to $N_{e^+} \sim 10^{49}$, and the intensity of the 511 
keV line would decrease by a factor of three. In Fig.~5, we show the 
changes of positron number, gas cloud density and the stregth of 511 keV 
flux as a function of time.

\begin{figure*}
\vskip 5mm
 \includegraphics[width=100mm]{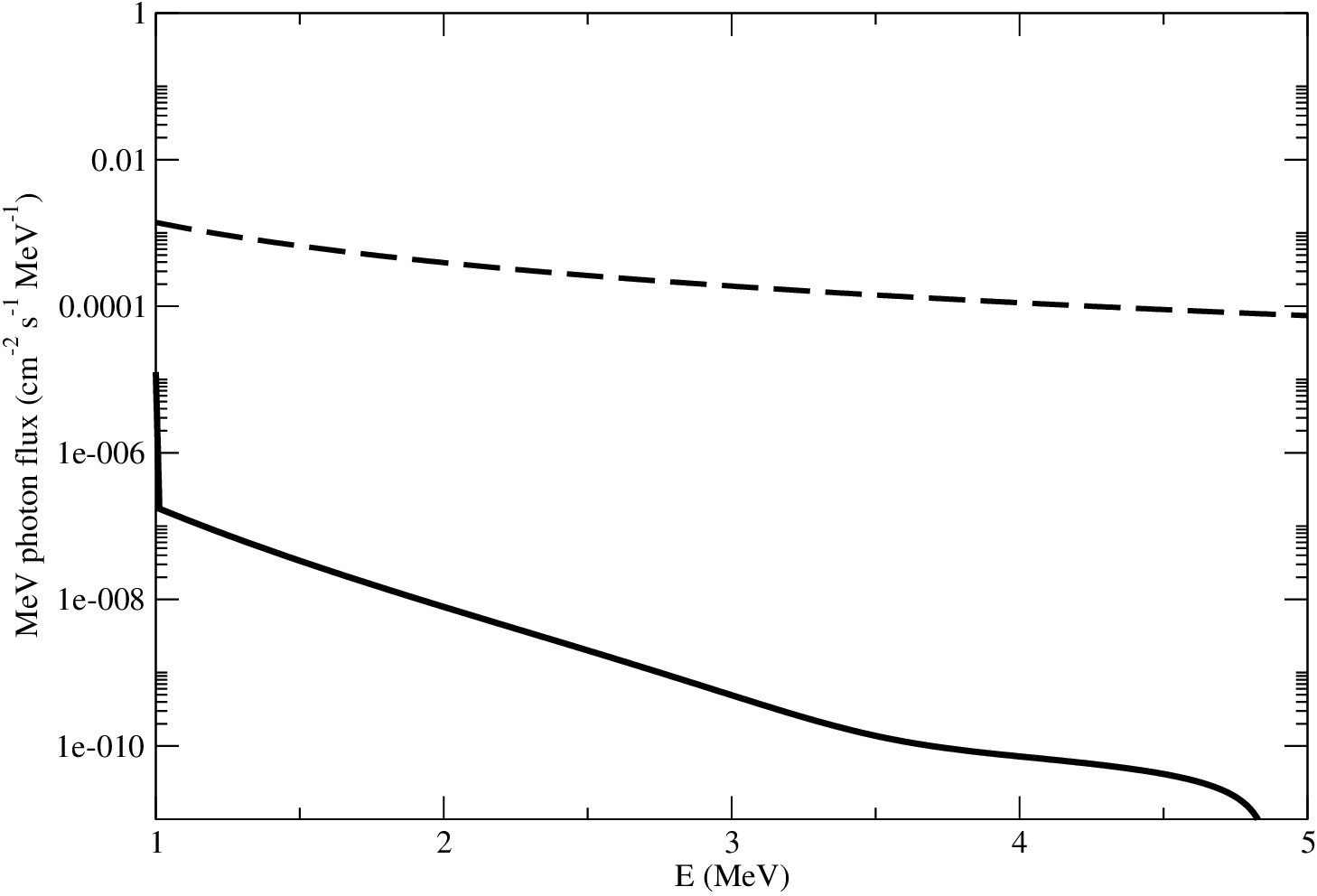}
 \caption{The solid and dashed lines represent the MeV photon spectrum 
due to positron-electron annihilation after passing through the dense 
cloud and the observed MeV photon spectrum respectively \citep{Beacom}.}
\vskip 5mm
\end{figure*}

\begin{figure*}
\vskip 5mm
 \includegraphics[width=100mm]{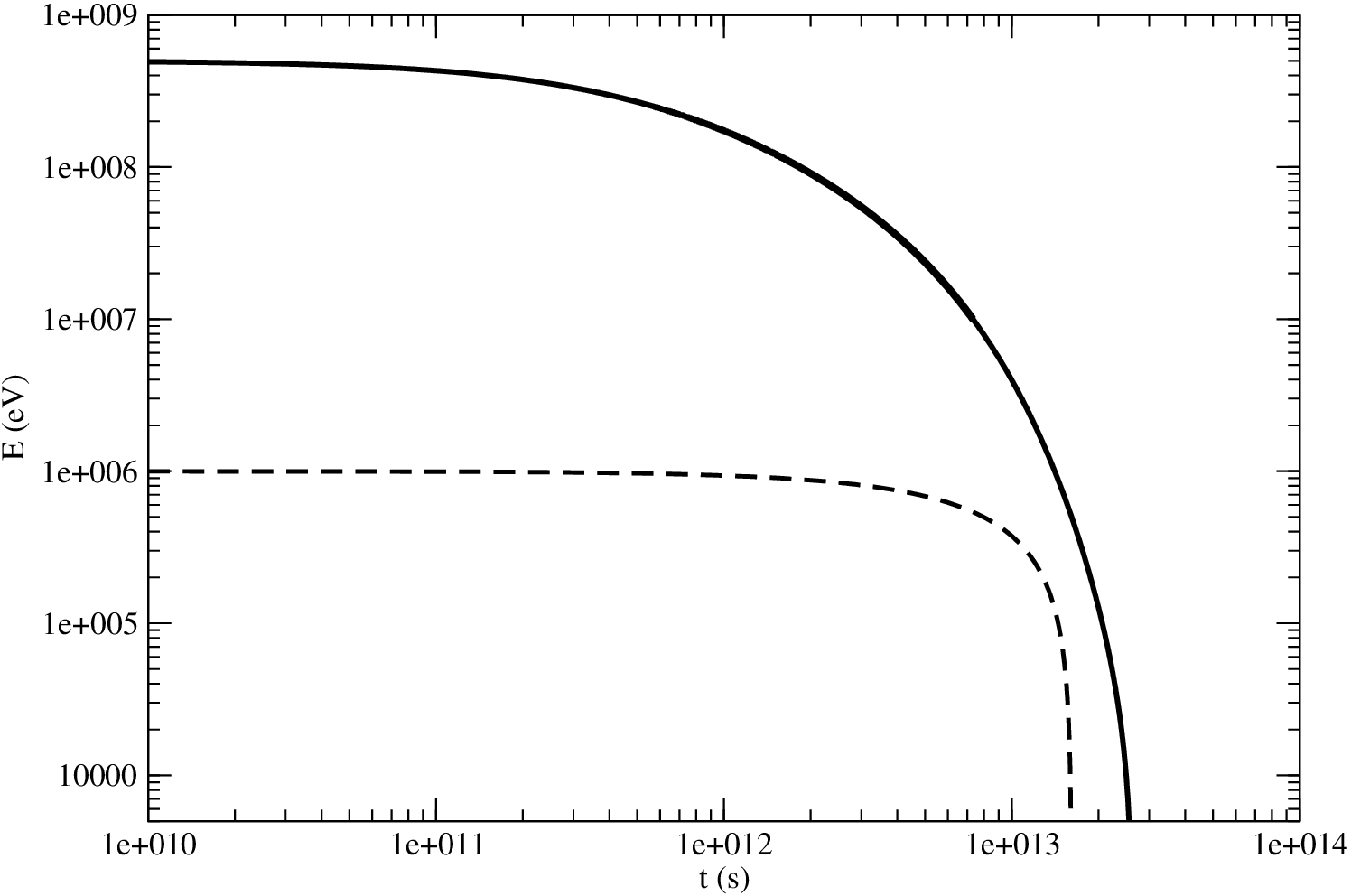}
 \caption{The cooling of a 1 MeV positron (dashed line) and a 0.5 GeV 
positron (solid line).} 
\vskip 5mm
\end{figure*}

\begin{figure*}
\vskip 5mm
 \includegraphics[width=100mm]{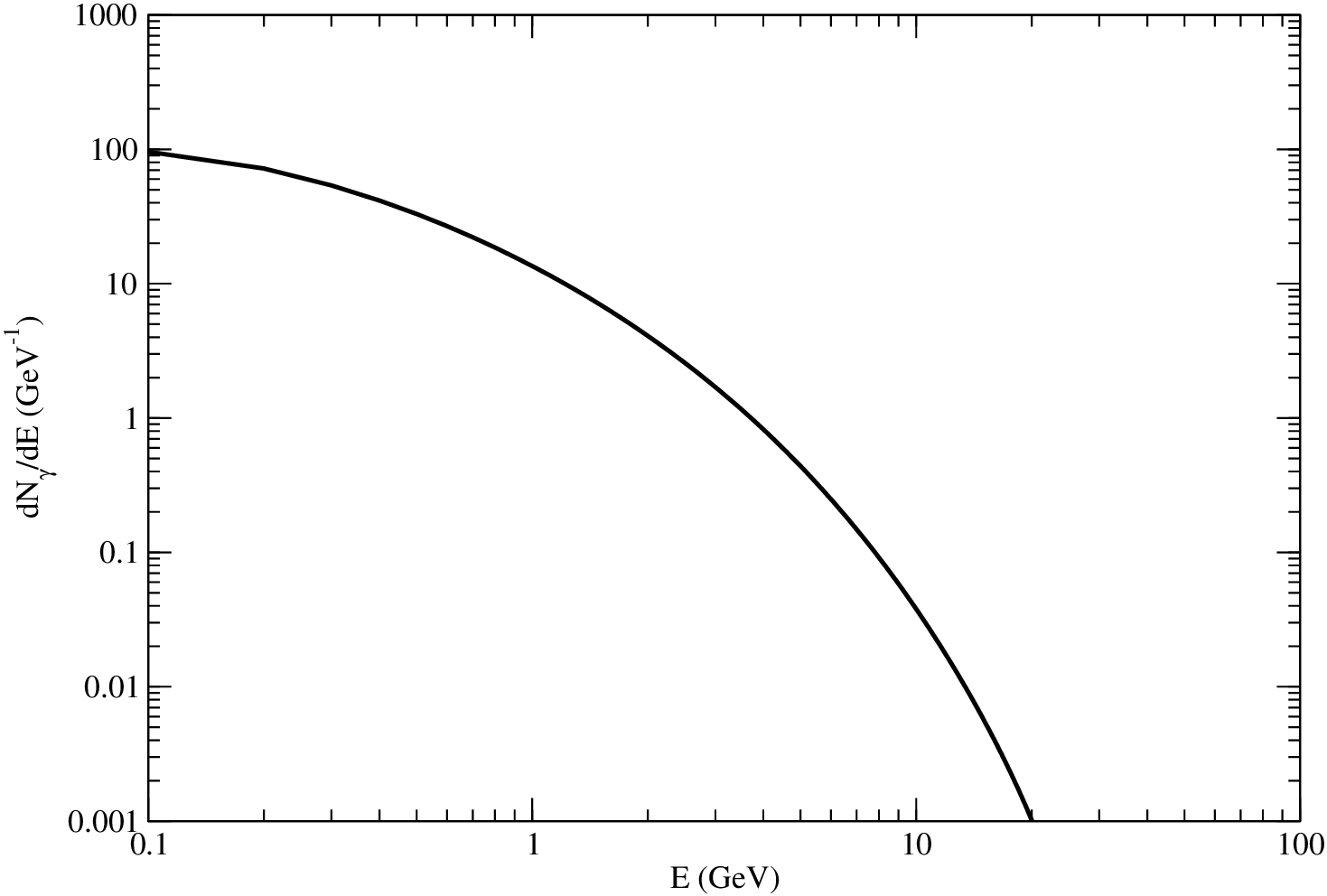}
 \caption{The photon spectrum per one dark matter annihilation 
\citep{Cembranos}. Here, we assume $m=40$ GeV.}
\vskip 5mm
\end{figure*}

\begin{figure*}
\vskip 5mm
 \includegraphics[width=100mm]{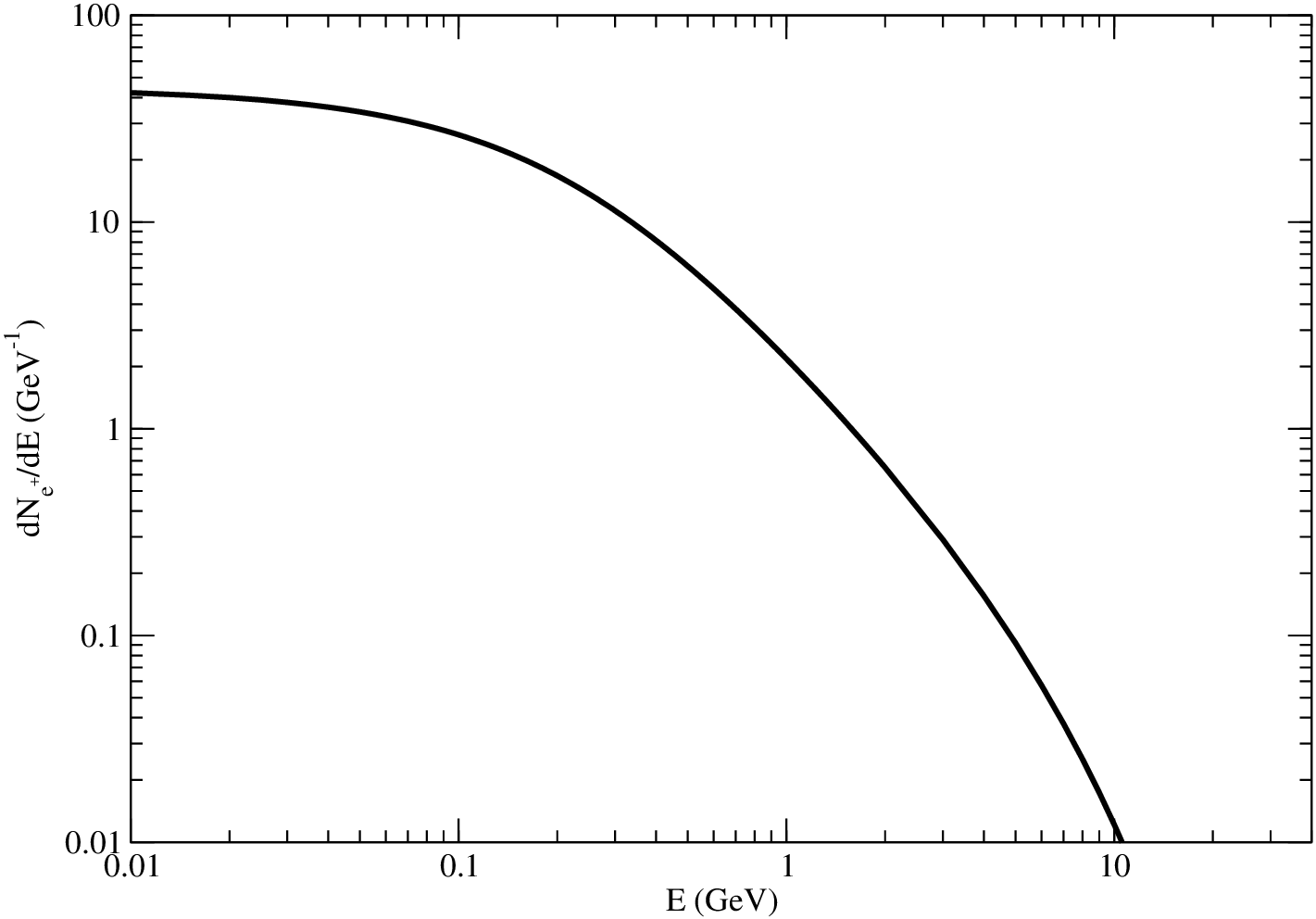}
 \caption{The positron spectrum per one dark matter annihilation 
\citep{Borriello}. Here, we assume $m=40$ GeV.}
\vskip 5mm
\end{figure*}

\begin{figure*}
\vskip 5mm
 \includegraphics[width=100mm]{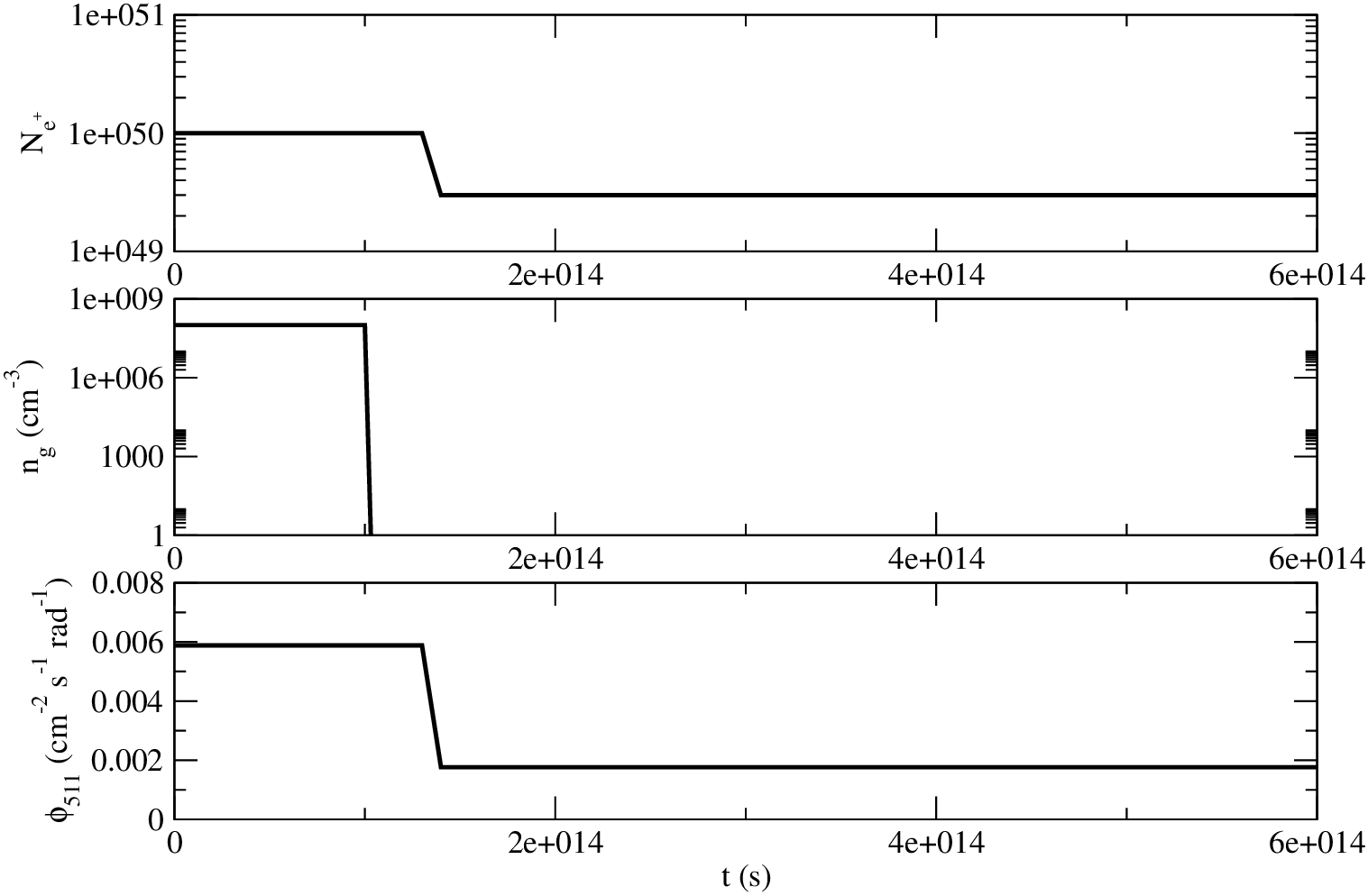}
 \caption{The changes of positron number, gas cloud density and the 
strength of 511 keV flux as a function of time. Here, we assume that the 
time of existence of the gas cloud $t_e$ and the cooling time scale for 
positrons $t_c$ are $10^{14}$ s and $2 \times 10^{13}$ s respectively.} 
\vskip 5mm
\end{figure*}

\section{Discussion and Conclusion}
In this letter, we discuss a possible model to explain the observed 511 
keV line. Following the dark matter annihilation scenario that used to 
account for the GeV gamma-ray excess in the Galactic Centre, the gamma-ray 
produced by the dark matter annihilation can generate enough positrons 
($\sim 10^{43}$ s$^{-1}$) through electron-positron pair-production inside 
the dense cloud, if we assume the existence of a dense 
gas cloud surrounding the supermassive black hole $10^{14}$ s ago. The 
cooling time scale (diffusion time scale) of positrons is $t_c \sim 
10^{13}$ s. Therefore, the 
strong 511 keV line observed is just a single event, which last for about 
$\sim 10^{13}$ s after the disappearance of the dense cloud. 

However, the uncertainties in the cloud size and density may significantly 
affect the optical depth for pair-production. Our model would fail if the 
optical depth near the supermassive black hole is too small for the 
high-energy gamma-ray. Therefore, more observational data from the 
parsec region near the supermassive black hole is needed to verify our 
proposed model. To conclude, by assuming a large dense cloud with number 
density $\sim 10^8$ cm$^{-3}$, it can generate enough positrons to account 
for the 511 keV line and explain the excess GeV gamma-ray observed in the 
Galactic Center.

\label{lastpage}

\end{document}